\documentclass[pre,
showpacs, twocolumn
]{revtex4}
\usepackage{graphicx,amsmath,amssymb,dcolumn,subfigure}

\begin{document}
\title{Dynamics in the smectic phase of stiff viral rods}

\author{Emilie Pouget,$^{1,2}$ Eric Grelet,$^{1,*}$ and M. Paul Lettinga$^{2,}$}
\altaffiliation{Corresponding authors: p.lettinga@fz-juelich.de; grelet@crpp-bordeaux.cnrs.fr}
\affiliation{$^1$ Universit\'{e} de Bordeaux, Centre de Recherche Paul-Pascal - CNRS,
 115 Avenue Schweitzer, 33600 Pessac, France\\
$^2$ IFF,
Institut Weiche Materie, Forschungszentrum J\"ulich, D-52425
J\"ulich, Germany}

\date{\today}

\begin{abstract}
We report on the dynamics in colloidal suspensions of stiff viral
rods, called fd-Y21M. This mutant filamentous virus exhibits a
persistence length three times and half bigger than the wild-type, fd-wt. Such
virus system can be used as a model system of rod-like particle to
study their self-diffusion. In this paper, the physical features such as rod contour length
and polydispersity have been determined for both viruses. The effect of viral rod flexibility on the location of the nematic-smectic
phase transition has been investigated, with a focus on the underlying
dynamics studied more specifically in the smectic phase. Direct visualization of the stiff fd-Y21M at the scale of a single particle
has shown the mass transport between adjacent smectic layers, as was found
earlier for the more flexible rods. We could relate this hindered diffusion
with the smectic ordering potentials for varying rod concentrations.
The self-diffusion within the layers is far more pronounced for the stiff
rods as compared to the more flexible fd-wt viral rod. 
\end{abstract}

\pacs{61.30.Hn, 61.30.Dk, 82.70.Dd}
 \maketitle

\section{Introduction}

Self-organization of Brownian particles is a subject of fundamental interest
since it deals with the most simple pathway to form structured materials. Anisotropic
particles with hard core interactions play an important role in this field because the shape anisotropy can lead to self-assembled materials with a hierarchy of ordering: the liquid crystal states. For these mesophases, one aims the connection between the macroscopic properties and the microscopic features of the
constituents. Onsager was the first to make such a connection in his seminal theoretical work \cite{Onsager}, showing how slender hard rods undergo a transition from the isotropic liquid phase 
to the nematic phase with a long ranged orientational order of the rods.
This work was expanded to the transition from the nematic to the smectic or lamellar phase, in which rods are organized in regular spaced layers, possessing still liquid ordering within the layer \cite{Mulder87,Wen87}. Besides this one dimensional positional order, simulations \cite{Frenkel88,Hess91,Lowen99,Bates99,Cifelli06,Cinacchi09} and experimental studies \cite{Wen89,Dogic01,Grelet08,Dogic06,GreletPRL08} revealed that transitions to more ordered states exhibiting two- and three dimensional long-ranged positional order can also be realized. For particles with hard core interactions these transitions are driven by entropy. Since this entropic gain is directly related with the accessible free volume, dynamics studies at the single particle level provides fundamental information for understanding the physical behavior of mesophases of rod-like particles.
In literature, only few studies where dynamical phenomena are tracked at the scale of the individual
anisotropic particles are known. The most versatile rod-like model systems have been sofar filamentous bacteriophages fd-wt \cite{Lettinga05,Lettinga10,Lettinga07,Grelet08}. These biological rods have an aspect ratio larger than 100 and each virus
is essentially identical to each other in diameter and contour
length, allowing for the formation of the smectic phase \cite{Dogic06}.
Recently, we have visualized, for the first time, the dynamics in
this smectic phase at the scale of the single particle
\cite{Lettinga07,Grelet08}. By tracking the individual viruses using
fluorescence microscopy it was found that self-diffusion mainly occurs via a jumping-type diffusion process, where viruses jump between adjacent smectic layers overcoming the interlayer energy barrier. In contrast, the diffusion within the smectic layers is much reduced. Thus after the nematic-smectic (N-Sm) phase transition, the diffusion anisotropy $\frac{D_\|}{D_\perp}$ increases, where $D_\|$ and perpendicular $D_{\perp}$ are the diffusion rate parallel and perpendicular to the long axis rods (or equivalently normal to the smectic layers), respectively .

This dynamical behavior of rods in the smectic phase was very surprising since the smectic structure is thought to be a regular stacking of slabs exhibiting two dimensional liquid-like order. Thus, it was expected that the rods would  diffuse within the layers rather than jump across the energy barrier that sets the layers.
Although simulations also display the jumping dynamics of the rods, they show that $D_\|$ and $D_{\perp}$ are in the same range of values \cite{Patti09,Matena10,Cinacchi09}. More precisely, experiments \cite{Hara85,Krueger82} and simulations \cite{Cinacchi09,Cifelli06,Zhang10} of thermotropic liquid crystals indicate that $D_{\perp}$ becomes faster than $D_\|$ shortly after the N-Sm transition, in this case with decreasing temperature. It should be noted, however, that the aspect ratio between the length $L$ and thickness $d$ of these systems was always at least one order of magnitude smaller than for fd-wt. In addition, it should be considered that fd-wt is relatively flexible.
The location of the different phases depends on the main features of the rods, such as the aspect ratio \cite{Graf99}, charge \cite{Wensink07} and flexibility \cite{Hidalgo05,Tkachenko96,vanderschoot96}. We therefore conjecture that such rod parameters could play a major role in the dynamics of filamentous virus suspensions.

Very recently Barry and coworkers reported the liquid-crystalline properties of the viral mutant fd-Y21M, which has a single point mutation in the amino-acid sequence of the major coat protein compared to fd-wt \cite{Barry09}. Due to this mutation the persistence length of the rod is 3.5 times larger than the persistence length of fd-wt virus, while the aspect ratio is assumed to be the same. The authors found that the increased stiffness caused the isotropic-nematic transition to be exactly located at the value predicted by Onsager's theory \cite{Onsager,Barry09}. Fd-Y21M is thus an ideal model system of rigid rod for studying the effect of stiffness on the phase behavior and on the underlying dynamics of rod-like particles.

In this paper we therefore compare the location of the N-Sm phase transition of fd-Y21M and fd-wt and determine the self-diffusion behavior of these viral rods. We first focu, on the specific features of fd-Y21M and its effect on the
location of the nematic-smectic phase transition. In the second part, we employ fluorescence video microscopy to study the self-diffusion in the lamellar mesophase. We show that the diffusion between the layers takes place by discrete jumps as for fd-wt, resulting in an diffusion anisotropy D$_\|$/D$_{\perp}\gg1$.
Contrary to fd-wt, this anisotropy decreases after the N-Sm phase transition due to the fact that the self-diffusion within the smectic layers is much more pronounced compared to the case of the more flexible viruses. Because the smectic range is found to be wider for stiff viruses, the effect of rod concentration on the dynamics of the smectic phase has been also studied.

\section{Experimental part}

Using standard biological procedures, large quantities of fd-Y21M and fd-wt bacterioviruses were grown and
purified using E. Coli XL1-Blue strain as bacterial host. In this study, the ionic strength has been fixed at I = 20 mM by a dialysis of both virus suspensions against a TRIS-HCl-NaCl buffer at pH
= 8.2. After
dialysis, the virus suspensions were ultracentrifuged at about 200
000 g during 3 hours. The supernatant was discarded and the viruses were
resuspended in the same buffer. We performed experiments on four
different concentrations of fd-Y21M: one in the nematic phase close
to the N-Sm transition ([N] = 89.9 $\pm$ 0.5 mg/mL) and three in the
smectic phase ([Sm1] = 93.8 $\pm$ 0.9 mg/mL, [Sm2] = 96.7 $\pm$ 1.0
mg/mL and [Sm3] = 98.9 $\pm$ 0.5 mg/mL). For fd-wt, two concentrations have been investigated:
one in the nematic phase close to N-Sm transition ([N] = 110 $\pm$ 2.0 mg/mL), and one in the narrow smectic range
([Sm] = 115 $\pm$ 2.0 mg/mL). Concentrations of the virus
suspensions were determined using absorption spectroscopy with an optical
density (OD) for a 1 mg/mL virus solution of OD$^{269nm}_{10mm}$ =
3.84 and 3.63 for of fd-wt and fd-Y21M, respectively
\cite{Barry09}. The smectic phase was evidenced by its iridescence
observed by illumination of the sample with white light, and confirmed by optical
microscopy using differential interference contrast (DIC).

Video fluorescence microscopy has been used to monitor the dynamics
of individual labeled colloidal rods in the background of the
nematic and smectic mesophases formed by identical but unlabeled
rods, where about one fd rod out of 10$^5$ has been labeled with the
dye Alexa-488 (Invitrogen), according to a previously published
protocol \cite{Lettinga05}. The colloidal scale of the fd virus
enables the imaging of individual rods by fluorescence microscopy,
as well as smectic layers by DIC microscopy \cite{Lettinga07}.
Individual fluorescently labeled viruses were visualized using an inverted
microscope (Zeiss Axiovert) equipped with a high numerical aperture oil
objective (100X PlanFluor NA 1.35) and a Mercury lamp. Images were
collected with a high sensitivity EM-CCD camera (Hamamatsu C9100) operating in a
conventional mode. Acquisition times were kept at minimum 51~ms in order to reduce blurring effects, while almost all the movies have a duration of about 25~s.
Usually, some loss of particles is evidenced after about 10~s of tracking due to both diffusion of particles out of the focus plan and photobleaching of the fluorescent dyes labeled on the viruses.

For the transmission electron microscopy (TEM) observations, diluted virus suspensions of about 10$^{-3}$ mg/mL
were deposited onto freshly glow-discharged
200-mesh Formvar/carbon-coated grids (purchased from Agar) and
allowed to settle for 1 min. Grids were then blotted, briefly rinsed
with distilled water and stained with 2\% uranyl acetate stainer 1
min. Grids were again blotted and rinsed with distilled water, and
observed with a Hitachi H-600 microscope operating at 75kV. Images
were recorded with a DVC type AMT (Advanced Microscopy Techniques) CCD camera.

\begin{figure}
\includegraphics[width=0.49\textwidth]{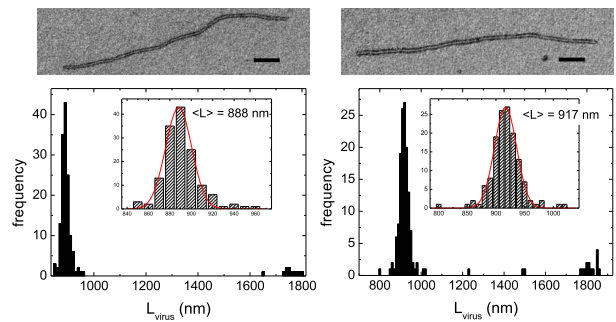}
\caption{\label{Fig1} (color online) Uranyl acetate negative stained virus images obtained by transmission electron microscopy
(The scale bars correspond to 100 nm) and distributions (done on 156 measurements for both viruses) of the virus contour length for fd-wt (left)
and fd-Y21M (right). The insets correspond to a focus on the
main distribution peak which has been fitted by a Gaussian function (red lines)
to provide the average value and the standard deviation of the virus contour length. }
\end{figure}

\begin{figure}
\includegraphics[width=0.4\textwidth]{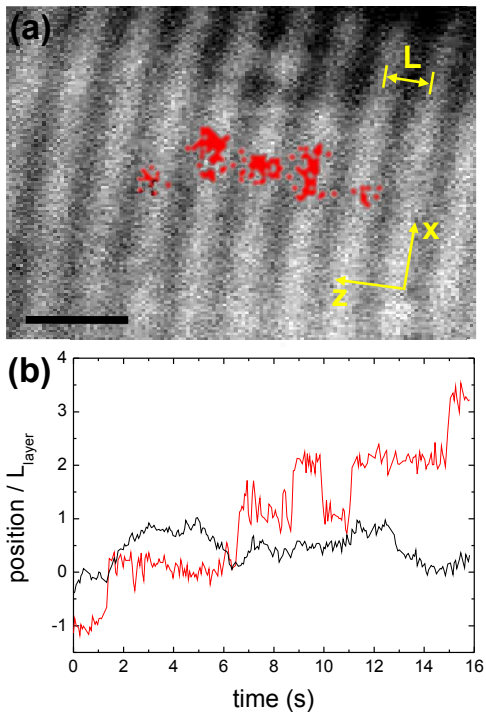}
\caption{\label{Fig2} (color online) (a) Overlay of differential interference
contrast (DIC) image and the trajectory of a virus tracked by
fluorescence microscopy in the smectic phase ([fd-Y21M] = 96.7
mg/mL). The trace shows clearly the jumps between adjacent smectic
layers and the self-diffusion occurring within the layers. The scale bar
represents 2 $\mu$m. (b) Displacement corresponding to the particle tracked
in (a), in the direction parallel (z, in red) and perpendicular (x, in black) to
the normal to the smectic layers. 
}
\end{figure}

\section{Results and discussion}

Although the persistence length of fd-Y21M has been measured to be 3.5 times higher than the one of fd-wt virus \cite{Barry09}, less is known about the polydispersity and contour length of fd-Y21M. From negative staining TEM images (Fig. \ref{Fig1}), we obtained a contour length distribution for fd-Y21M which is centered around $<L> = 917$~nm whereas a similar procedure applied to fd-wt gives an average value of $<L> = 888$~nm. This value for fd-wt is in agreement with the previous published value,  which is $<L> = 0.88 \pm 0.03~\mu m$ \cite{Day88}. The amino-acid mutation of the coat proteins does not only influence the rigidity of the rod but also its contour length, by an increase of about 3\%. Even if the virus contour length is mainly fixed by the size of its DNA \cite{Marvin06}, which is expected to be similar for both viruses, it has been reported that the amino-acid mutation in fd-Y21M significantly affects the symmetry of the coat protein arrangement \cite{Tan99}, which could therefore be at the origin of the slight change observed in the contour length \cite{Hunter87}. The use of XL1-Blue host as E. Coli strain for the growth of the bacteriophages, is supposed to produce highly monodisperse filamentous viruses \cite{Chapfd}. The polydispersity, $\sigma_L$, is defined by the relative standard deviation of the Gaussian distribution: $\sigma_L=(<L^2>-<L>^2)^{1/2}/<L>$, where the brackets indicate a statistical average. The numerical fit of the main peak of the virus contour length distribution (Fig. \ref{Fig1}) gives $\sigma_L^{fd-Y21M}$=2.1\% and $\sigma_L^{fd-wt}$=1.4\%. Note that these two values have not been deconvoluted from the measurement error, which is estimated to be around 1\%. Consequently, we consider no significant difference (including the error bars) in the measured polydispersity of both viruses.
Furthermore, the existence of longer viruses with a contour length centered on 1820 nm and 1764 nm for fd-Y21M and fd-wt respectively (Fig. \ref{Fig1}) is noteworthy. These lengths correspond to the
formation of viral dimers. 
These double-length particles contain two phage genomes, i.e. two single stranded DNA molecules \cite{Russel91}, and their probability of presence is around 1/10 compared to normal viral form (Fig. \ref{Fig1}).

The weak polydispersity of the both viruses allows for the formation of the smectic phase, which is certainly not destabilized by the presence of viral dimers which are easily accommodated in the layer spacing. Interestingly, the nematic-smectic phase transition of the stiff fd-Y21M occurs at lower concentration compared to fd-wt, which is consistent with the phase behavior observed at the isotropic-nematic phase transition \cite{Barry09} and with theoretical predictions\cite{vanderschoot96,Hidalgo05}.
Moreover, the smectic phase is obtained with a larger concentration range than fd-wt \cite{Grelet11}.
Using DIC microscopy imaging, we confirmed earlier findings \cite{Dogic97} that the smectic layer spacing of fd-wt is 0.89 $\mu$m, i.e. equal to the contour length of the virus. In contrast, a clear
difference is observed between the fd-Y21M length (L$_{virus}$ =
0.92 $\mu$m) and the interlayer distance (L$_{layer}$ = 0.96
$\mu$m), over the full concentration range studied here. This increase of the smectic layer spacing with increasing rod stiffness is in qualitative agreement with theoretical predictions \cite{Tkachenko96,vanderschoot96}.

In the following we will study to what extend the dynamics of the rods is affected by the higher rigidity. This influence can be directly observed in Fig. \ref{Fig2}a, which shows the overlay of the DIC image of the smectic phase of fd-Y21M and the trace of a
labeled virus tracked by fluorescence microscopy. This trace and the corresponding trajectory in Fig. \ref{Fig2}b display the diffusion of the particle through the smectic layers in quasi-quantized steps of one layer spacing, and also a clear diffusion within the layer, which was not observed for fd-wt \cite{Lettinga07,Grelet08}.

\begin{figure}
\includegraphics[width=0.45\textwidth]{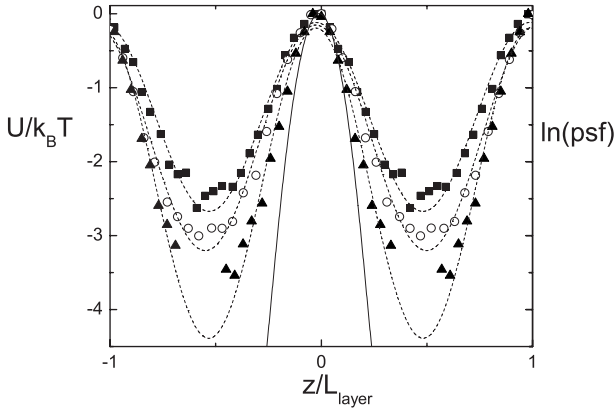}
\caption{\label{Fig_potential} Ordering potential (raw data) for the three different smectic concentrations corresponding to square, circle, and triangle symbols by increasing rod concentration, respectively. The dashed lines are fits to a sinusoidal potential. The solid line is the logarithm of the function $ln(psf)$ representing the smearing along the rod long axis. }
\end{figure}

\begin{figure}
\includegraphics[width=0.45\textwidth]{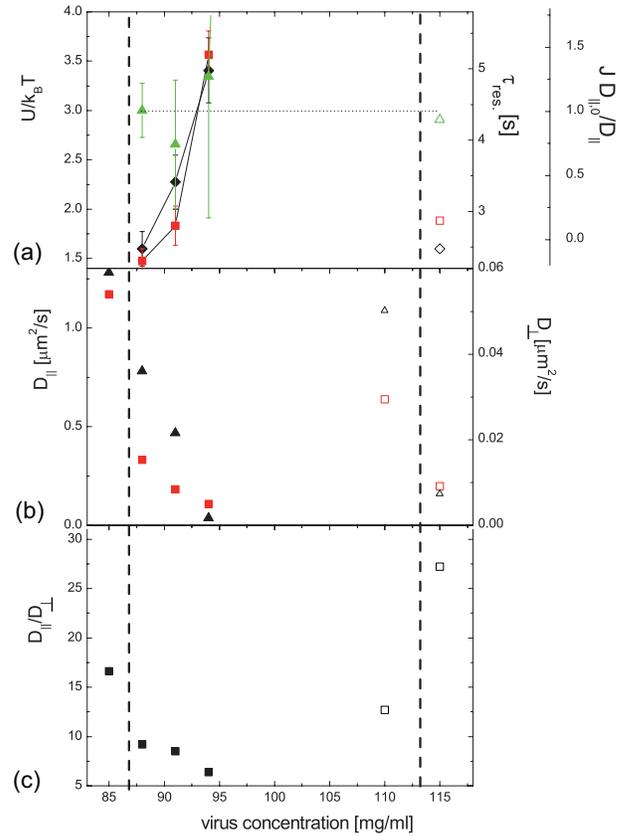}
\caption{\label{Fig_conc} (color online) Concentration dependence of the (a) amplitude of the final smectic ordering potential obtained after the deconvolution process (diamonds; see main text), virus residence time (squares) and barrier jumping scaling times $D_{N}/D_{Sm}$ (triangles); (b) diffusion rates parallel (squares) and perpendicular (triangles) to the normal to the smectic layers; (c) ratio of diffusion parallel and perpendicular to the normal to the smectic layers. The solid symbols represent the results for fd-Y21M while the open symbols represent those for fd-wt. The dashed lines at low and high concentration indicate the N-Sm phase transition for both fd-Y21M and fd-wt, respectively. }
\end{figure}

We first want to determine the concentration dependence of the ordering potentials in the smectic phase. Following the method used in our previous work \cite{Lettinga07}, this is done by converting the probability function $P(z)$ of finding a particle at position $z$ with respect to the middle of a layer to the smectic ordering potential $U_{layer}(z)$ via the Boltzmann's law $P(z)\sim \exp(-U_{layer}(z)/k_BT)$. To obtain the total ordering potential, the particle distributions in a single layer is added periodically to itself at all integer numbers of layer spacing $L_{layer}$ \cite{Lettinga07}. All potentials can be best fitted with a sinusoidal function $U_{layer}(z)=U_0 \sin(2\pi z/L_{layer})$, as shown in Fig. \ref{Fig_potential} by the dashed lines. Note that the distribution of the particles is a convolution of the real distribution with a function representing the smearing of the particle location due to the limited experimental resolution (solid line in Fig. \ref{Fig_potential}). Such function, the point spread function (psf), has been obtained with immobile viruses fixed on the glass cover slip of the sample under the same experimental conditions. The real distribution is found by a deconvolution process, where we obtain the test function for which the difference between the exponential of our experimentally obtained potential and the convolution of the smearing function with this test function is at its minimum. Our final potential is then given by the logarithm of this test function, whose amplitude is shown in Fig. \ref{Fig_conc}a.

The effect of the self-organization in smectic layers on the diffusion of the rods can be characterized in several ways. Since the jumping behavior dominates the dynamics, we first measure the average residence time, i.e. the time a rod-like particle stays within a given smectic layer without jumping during the duration of the experiment. 
As can be seen in Fig. \ref{Fig_conc}a, the concentration dependence of the residence time matches the one of the smectic ordering potential. With somewhat higher concentrations the residence time seems to effectively diverge to infinity.
%
%


\begin{figure}
\includegraphics[width=0.51\textwidth]{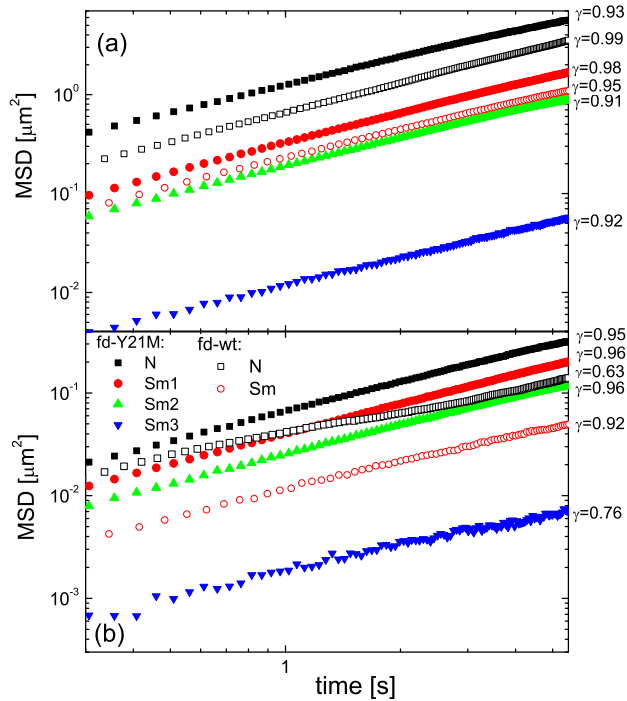}
\caption{\label{Fig_gamma} (color online) Log-log representation of the mean
square displacement (MSD) parallel (a) and perpendicular (b) to the normal to the smectic layers.
The solid and open symbols represent the results for fd-Y21M and fd-wt, respectively.
The solid lines represent the numerical fits
by a power law; MSD$\propto t^\gamma$. }
\end{figure}

To explore the effect of the smectic ordering potential on the diffusion, the mean
square displacements (MSD) of the virus parallel and perpendicular to
the normal to the smectic layers have been investigated, as shown in Fig. \ref{Fig_gamma}. The time evolution of the MSD given by $<r_i^2(t)>\;=\;D_i.t^\gamma$ provides the diffusion exponent $\gamma$ and the diffusion rate $D_i$ in the $i$ direction. When $\gamma \simeq$ 1, the self-diffusion has a Brownian (diffusive) behavior, while $\gamma < 1$ is characteristic of a sub-diffusive regime.
The stiff fd-Y21M rods display diffusive behavior when taking into account only the first five seconds. Only for the highest concentration the perpendicular diffusion becomes subdiffusive, as shown in Fig. \ref{Fig_gamma}. For fd-wt, the motion perpendicular to the long axis is subdiffusive and becomes more diffuse after the phase transition to the smectic phase, which is contrary to earlier statements that the system is subdiffusive within the layer for fd-wt  \cite{Lettinga07}. From our data, the only signature we found for the increase of the free accessible volume at the N-Sm transition, is the increase of diffusivity in the exponent $\gamma$ (from 0.63 to 0.92, see Fig. \ref{Fig_gamma}) for fd-wt system. Since the smectic range is broader for the stiff rods, we can nicely observe that the diffusivity within the layer decreases for higher concentrations to $\gamma=$0.76, approaching 0.5 which can be interpreted as a dynamic signature for the proximity of a transition to a glassy state.

It should be noted that the dynamics becomes subdiffusive for longer times (data not shown). We suspect this subdiffusive behavior to originate from the finite duration of acquisition movies and loss of tracked particles with time due to both diffusion out of the focus plan and dye photobleaching (See Experimental part). However, to discriminate whether or not such subdiffusive behavior is physical, specific studies of long time diffusion has to be performed using different experimental techniques such as fast recovery after photobleaching (FRAP).

Although the diffusion within the smectic layer is substantial for the stiff viral rods, the diffusion through the layers can still be regarded as the diffusion of a Brownian particle in a periodic potential. The range of concentrations in the smectic phase supplies a range of ordering potentials for which the theoretical predictions can be tested \cite{Festa78}:

\begin{eqnarray}\label{Eq_Msd}
D_{\|}=\frac{D_{\|,0}}{\langle e^{-U_{layer}(z)/k_BT}\rangle \langle e^{U_{layer}(z)/k_BT}\rangle}=J D_{\|,0}.
\end{eqnarray}

The brackets indicate here the average over one period of the smectic ordering
potential. The diffusion coefficient in the smectic phase can then
be calculated taking $D_{\|,0}$ as the diffusion coefficient in the
nematic phase close to the N-Sm phase transition, and using $U_{layer}$ as
plotted in Fig. \ref{Fig_conc}a. If the correction $J$ of the diffusion $D_{\|,0}$ due to the fact that the particles have to cross the potential barriers is valid, than we should find that $J\frac{D_{\|,0}}{D_\|}= 1$. Indeed, we find a good agreement between the measured potentials and parallel diffusion rates, using this approach since $J\frac{D_{\|,0}}{D_\|}\simeq 1$ for all concentrations, see triangles in Fig. \ref{Fig_conc}a and the dotted line as guide for the eye.

The diffusion coefficients parallel and perpendicular to the normal to the smectic layers, as summarized  in Fig. \ref{Fig_conc}b, enclose many physical informations on the dynamical origin of the location of the N-Sm phase transition and its dependence on the flexibility. In particular, we notice two important differences between fd-wt and fd-Y21M.

First, comparable diffusion rates are observed for both stiff and flexible rods just before the N-Sm transition, despite of the fact that this takes place at much higher concentrations for the flexible rods (Fig. \ref{Fig_conc}b). This nicely illustrates that the positional entropy of the flexible rods is higher in the nematic phase and therefore the phase transition is located at higher concentrations.

Second, after the transition to the smectic phase, the ratio in the diffusion parallel and perpendicular to the normal to the smectic layers,$\frac{D_\|}{D_\perp}$ as plotted in Fig. \ref{Fig_conc}c, \emph{decreases} in the case of the stiff fd-Y21M, while it \emph{increases} in the case of fd-wt \cite{Lettinga07}. Although the ratio of parallel over perpendicular diffusion is still higher than one, contrary to what is observed for thermotropic liquid crystal \cite{Hara85,Krueger82} and simulations\cite{Patti09,Matena10,Cinacchi09}, this behavior corresponds to what is expected for a smectic A phase formed of two-dimensional liquid layers.

Note that it is mainly the perpendicular diffusion, i.e. the diffusion within the layer, which is reduced dramatically after the N-Sm phase transition for the flexible rods as compared to the stiff rods. It drops to an equivalent value as observed for fd-Y21M, but within a much smaller concentration range. At only somewhat higher concentrations the system will undergo a phase transition to the columnar phase \cite{Grelet08}. This very low diffusion rate within the layer of flexible rods indeed hints that the columnar phase is proximate. This seems to be a dynamical signature for the fact that the smectic region is vanishing when increasing the rod flexibility. The result that the smectic layers are more densely pact for fd-wt compared to fd-Y21M could be also a cause of the low parallel diffusion rate, since rods will always partly penetrate neighboring layers, thus being frustrated in their diffusive motion by particles in the host layer as well as the neighboring layers.

Flexibility cannot not solely explain however why the diffusion ratio is much higher than one even for stiff rods. Here it is important to note that most simulations and thermotropic systems deal with particles with small aspect ratios as compared to virus system such as fd-Y21M. With the biochemical techniques available nowadays, the variation of both stiffness and aspect ration should be feasible, in particular to test if the smectic region completely disappears with increasing rod flexibility.

\section{Conclusions}

The effect of flexibility on the location of the N-Sm transition and the underlying dynamics was investigated. We show that, according to theoretical predictions, the phase transition occurs at higher concentrations when the rod becomes more flexible and that the layer spacing $L_{layer}$ is increased for stiffer rods. The main difference we found between the two viral systems is that the diffusion within the layer is much
 more reduced after the N-Sm transition for the flexible rods compared to the stiffer ones. This suggests that the columnar phase, where perpendicular motion should be quenched due to two-dimensional positional order, is favored as the rod flexibility increases. The fact that even for the stiffer particles $\frac{D_\|}{D_\perp}\gg 1$ in contrast with simulations and diffusion in thermotropic liquid crystals, is probably connected to the very high rod aspect ratio of the viral particles. Especially when the diffusion is scaled with the diameter of the rod, it appears that the diffusion within the layer is very fast and liquid-like (diffusive). For higher concentrations the dynamics changes and seems to become frozen: such dynamical behavior  is currently under investigations for these highly dense states.

\begin{acknowledgments}
This project was supported by the European network of excellence
SoftComp. We would like to thank Z. Dogic for providing us the strain of the mutant virus fd-Y21M.
\end{acknowledgments}


\begin{thebibliography}{99}

\bibitem{Onsager} L. Onsager, Ann. N.Y. Acad. Sci {\bf 51}, 627 (1949).

\bibitem{Mulder87} B. Mulder, Phys. Rev. A {\bf 35}, 3059 (1987).

\bibitem {Wen87} X. Wen and R.~B. Meyer,
  Phys. Rev. Lett. {\bf 59}, 1325(1987).

\bibitem{Frenkel88} D. Frenkel, H. N. W. Lekkerkerker and S. Stroobants, 
  Nature {\bf 332}, 822 (1988).
\bibitem{Hess91} S. Hess, D. Frenkel and M. P. Allen, 
  Molecular Physics {\bf 74}, 765 (1991).

  \bibitem{Lowen99}H. L\"owen, 
  Phys. Rev. E {\bf 59}, 1989 (1999).
  \bibitem{Bates99} M. A. Bates and G. R. Luckhurst, 
  J. Chem. Phys. {\bf 120}, 394 (2004).
  \bibitem{Cifelli06} M. Cifelli, G. Cinacchi and L. De Gaetani, 
  J. Chem. Phys. {\bf 125}, 164912 (2006).
  \bibitem{Cinacchi09} G. Cinacchi and L. De Gaetani, 
  Phys. Rev. E {\bf 79}, 011706 (2009).
  \bibitem{Wen89} X. Wen, R. B. Meyer and D. L. D. Caspar, 
  Phys. Rev. Lett. {\bf 63}, 2760 (1989).
  \bibitem{Dogic01} Z. Dogic and S. Fraden, 
  Phil. Trans. R. Soc. Lond. A {\bf 359}, 997 (2001).

\bibitem{Grelet08} E. Grelet, M. P. Lettinga, M. Bier, R. van Roij and P. van der Schoot, 
  J. Phys.: Condens. Matter {\bf 20}, 494213 (2008).

\bibitem{Dogic06} Z. Dogic and S. Fraden, 
  Current Opinion in Colloid \& Interface Science {\bf 11}, 47 (2006).
\bibitem{GreletPRL08} E. Grelet, Phys. Rev. Lett. {\bf 100}, 168301 (2008).
 \bibitem{Lettinga05} M. P. Lettinga, E. Barry, Z. Dogic, 
  Europhys. Lett. {\bf 71}, 692 (2005).
\bibitem{Lettinga07} M. P. Lettinga and E. Grelet, 
  Phys. Rev. Lett. {\bf 99}, 197802 (2007).
\bibitem{Lettinga10} M. P. Lettinga, Z. Zhang, J. K .G. Dhont, S. Messlinger and G. Gompper,
Soft Matter {\bf 6}, 4556 (2010).
\bibitem{Matena10} R. Matena, M. Dijkstra and A. Patti, Phys. Rev. E {\bf 81} 021704 (2010). \bibitem{Patti09} A. Patti, D. El Masri, M. Dijkstra and R. van Roij,
Phys. Rev. Lett. {\bf 103}, 248304 (2009).
\bibitem{Hara85} M. Hara, H. Tenmei, S. Ichikawa, H. Takezoe and A. Fukuda
Jap. J. App. Phys. {\bf 24}, L777 (1985),

\bibitem{Krueger82} G.~J. K{r\"u}ger, Physics Reports {\bf 82}, 229 (1982).
\bibitem{Zhang10} Z. Zhang and H. Guo, J. Chem. Phys. {\bf 133} 144911 (2010).
\bibitem{Graf99} H.~Graf, H.~L\"owen, Phys. Rev. E {\bf 59}, 1932 (1999).
\bibitem{Wensink07} H. H. Wensink, J. Chem. Phys. {\bf 126}, 194901 (2007).
\bibitem{Hidalgo05} R. C. Hidalgo, D. E. Sullivan and J. Z. Y. Chen, Phys. Rev. E {\bf 71}, 041804 (2005).
\bibitem{Tkachenko96} A. V. Tkachenko, Phys. Rev. Lett. {\bf 77}, 4218 (1996).
\bibitem{vanderschoot96} P. van der Schoot, J. Phys. II France {\bf 6}, 1557 (1996).\bibitem{Barry09} E. Barry, D. Beller and Z. Dogic, Soft Matter {\bf 5}, 2563 (2009).
\bibitem{Day88} L. A. Day, C. J. Marzec, S. A. Reisberg, A.Casadevall, Annu. Rev. Biophys. Biophys. Chem.
 {\bf 17}, 509 (1988); L. C. Welsh, M. F. Symmons, C. Nave, R. N. Perham, E. A. Marseglia, and D. A. Marvin, Macromolecules {\bf 29}, 7075 (1996).
\bibitem{Marvin06} D. A. Marvin {\it et al.}, J. Mol. Biol. {\bf 355}, 294 (2006).
\bibitem{Tan99} W. M. Tan {\it et al.}, 
J. Mol. Biol. {\bf 286}, 787 (1999).
\bibitem{Hunter87} G. J. Hunter, D. H. Rowitch, and R. N. Perham, Nature {\bf 327}, 252 (1987).
\bibitem{Chapfd} Z. Dogic and S. Fraden, in {\it{Soft Matter, volume 2}}, edited by G. Gompper and M. Schick
(WILEY-VCH, Weinheim, 2006)
\bibitem{Russel91} M. Russel, Molecular Microbiology {\bf 5}, 1607 (1991).
\bibitem{Grelet11} E. Grelet {\it et al.}, unpublished results.
\bibitem{Dogic97} Z. Dogic and S. Fraden, Phys. Rev. Lett.  {\bf 78}, 2417 (1997).
\bibitem{Festa78} R. Festa and E. Galleani d'Agliano, Physica, {\bf 90a}, 229 (1978).




\end{thebibliography}
\end{document}